\documentclass[preprint]{elsarticle}

\usepackage{graphicx}
\usepackage{amsmath,amssymb}

\usepackage[english]{babel}
\usepackage{graphics}
\usepackage{color}
\usepackage{wrapfig}
 \usepackage[usenames,dvipsnames]{pstricks}
 \usepackage{epsfig}
 \usepackage{pst-grad} 
 \usepackage{pst-plot} 
\usepackage{times}

\def\f{\frac}
\def\be{\begin{equation}}
\def\ee{\end{equation}}
\def\ba{\begin{eqnarray}}
\def\ea{\end{eqnarray}}

\def\scri{\mathcal{I}}

\def\s1pr{\mathcal{I}^{1+}_{\rm R}}

\def\spr{\mathcal{I}^{+}_{\rm R}}

\def\ub{\underbar}

\def\Dp{\partial_{+}}
\def\Dm{\partial_{-}}
\def\dd{{\rm d}}

\def\k{\kappa}

\def\y{{y}}

\begin{document}

\title{Two-dimensional quantum black holes: Numerical methods}
    \author[rvt]{Fethi M Ramazano\u{g}lu\corref{cor}}    
    \ead{framazan@princeton.edu}
    \author[rvt]{Frans Pretorius}
    \ead{fpretori@princeton.edu}
    \address[rvt]{Department of Physics, Princeton University, Princeton, NJ, 08544, USA }
\cortext[cor]{Corresponding author: Jadwin Hall, Princeton, NJ 08544, USA, Phone: 609-258-4355, Fax: 609-258-1124 }

\begin{abstract}

{\small We present details of a new numerical code designed to study the formation and evaporation of 2-dimensional
black holes within the CGHS model. We explain several elements of the scheme that are crucial to resolve the
late-time behavior of the spacetime, including regularization of the field variables, compactification of the 
coordinates, the algebraic form of the discretized equations of motion, and the use of a modified Richardson
extrapolation scheme to achieve high-order convergence. Physical interpretation of our results will be discussed in detail elsewhere.}
\end{abstract}


\maketitle
\section{Introduction}\label{sec_intro}

The Callan-Giddings-Harvey-Strominger (CGHS) model \cite{cghs} is a two-dimensional model of quantum gravity which has 
attracted attention due to the fact that it has black hole solutions with many of the qualitative features of four-dimensional 
black holes, while being technically easier to investigate. Various properties of black holes in this model, and other models 
inspired by it, have been studied extensively using analytical and numerical methods \cite{cghs_other,PiranStrominger,Lowe}; 
for pedagogical reviews see \cite{reviews}. A recent focus point has been on using the CGHS model to investigate the black hole 
information loss problem \cite{ATV}, where the
importance of understanding the asymptotic behavior of the fields near right-future null infinity $\scri_R^+$ was emphasized.
In particular, sufficient conditions for the unitarity of the S-matrix were given.
Although the full quantum equations are too complicated to solve, in the mean field approximation (MFA) the model
reduces to a coupled set of non-linear partial differential equations, possessing a well-posed characteristic initial
value formulation. Unfortunately, even for these equations, analytical solutions are not known except in special limiting cases. 
Therefore, to explore black hole formation and evaporation, numerical methods are essential. 

In this paper, we give details of the methods we have devised for accurate numerical calculations of the fields 
and related physical quantities in the CGHS model. We give special attention to the macroscopic mass limit, 
which is the most challenging case to calculate and which has not been properly investigated before. An outline 
of the rest of the paper is as follows.
In Sec.~\ref{sec_model} we introduce the CGHS model, describe the variable definitions and 
conventions we use (which closely follows~\cite{ATV}), the analytical equations that we
discretize, and the initial data we use.
In Sec.~\ref{sec_prep}, we describe some of the issues that would cause naive
discretization of the equations to fail to uncover the full spacetime, and
how to overcome them; this includes regularization of otherwise asymptotically-divergent
field variables, compactification of the coordinates, the particular discretization scheme, and
use of Richardson extrapolation ideas to increase the accuracy of the solution.
In Sec.~\ref{sec_prep} we also discuss setting initial conditions near $\scri$, and
how we extract the desired asymptotic properties of the solution.
In Sec.~\ref{sec_tests} we describe various tests to demonstrate we have a
stable, convergent numerical scheme to solve the CGHS equations. 
We summarize and conclude in Sec.~\ref{sec_conc}.

\section{CGHS Model}\label{sec_model}
The action of the 2-dimensional (2D) CGHS model is given by
\ba S(g,\phi,f) = \f{1}{G}\textstyle{\int}\!\! &\dd^2V&\!\!
e^{-2\phi}\, \left(R + 4 g^{ab}\nabla_a\phi \nabla_b \phi + 4
\k^2\right)
\nonumber\\
&-& \f{1}{2} \textstyle{\int} \dd^2V\, g^{ab}\nabla_a f \nabla_b f \ .
\ea
where $g^{ab}$ is the metric, $R$ is the Ricci scalar, $\phi$ is a dilaton field, $f$ is a massless
scalar field, $G$ is Newton's constant and $\kappa$ is a constant of dimension inverse-length. 
Note that this action is {\em similar} to though {\em not} exactly the same as what would
be obtained by dimensional reduction of the 4D Einstein-Klein-Gordon equations in spherical symmetry.

\begin{figure}
\psset{unit=1cm}
\begin{pspicture}(0,-5.365)(12.989688,5.365)
\psline[linewidth=0.04,arrowsize=0.05291667 2.0,arrowlength=1.4,arrowinset=0.4]{-<<}(4.2555614,1.7026116)(5.6972275,0.31639418)
\psline[linewidth=0.04,arrowsize=0.05291667 2.0,arrowlength=1.4,arrowinset=0.4]{-<<}(5.6972275,0.31639418)(7.1388936,-1.0698233)
\psline[linewidth=0.04,arrowsize=0.05291667 2.0,arrowlength=1.4,arrowinset=0.4]{-<<}(7.1388936,-1.0698233)(8.6,-2.525)
\psbezier[linewidth=0.072,linestyle=dashed,dash=0.16 0.16]{cc-}(7.7765536,2.4788933)(6.967927,1.6379215)(6.979552,-0.91660976)(7.5713935,-1.4856884)
\psbezier[linewidth=0.04]{cc-}(4.111395,1.8412334)(4.688061,1.2867464)(7.1824603,1.8610364)(7.7765536,2.4788933)
\psline[linewidth=0.06,arrowsize=0.05291667 2.0,arrowlength=1.4,arrowinset=0.4]{cc->}(7.7765536,2.4788933)(8.4,3.075)
\psline[linewidth=0.06,arrowsize=0.05291667 2.0,arrowlength=1.4,arrowinset=0.4]{cc->}(8.170239,-3.1713288)(9.140592,-2.1621625)
\psline[linewidth=0.06,arrowsize=0.05291667 2.0,arrowlength=1.4,arrowinset=0.4]{cc->}(3.2186708,-3.1269698)(2.353671,-2.2952394)
\psline[linewidth=0.04,arrowsize=0.05291667 2.0,arrowlength=1.4,arrowinset=0.4]{-<<}(2.8138952,3.088829)(4.2555614,1.7026116)
\usefont{T1}{ptm}{m}{it}
\rput(10.51406,1.8){\large $\scri^+_R$}
\usefont{T1}{ptm}{m}{it}
\rput(9.011406,-4.02){\large $\scri^-_R$}
\usefont{T1}{ptm}{m}{it}
\rput(0.8, 1.8){\large $\scri^+_L$}
\usefont{T1}{ptm}{m}{it}
\rput(3.0014062,-4.02){\large $\scri^-_L$}
\usefont{T1}{ptm}{m}{it}
\rput(2.3814063,-2.82){$z^-$}
\usefont{T1}{ptm}{m}{it}
\rput(9.221406,-2.82){$z^+$}
\usefont{T1}{ptm}{m}{it}
\rput(3.3314064,-0.82){flat metric}
\psline[linewidth=0.06,arrowsize=0.05291667 2.0,arrowlength=1.4,arrowinset=0.4]{cc->}(4.0,4.875)(5.8,1.875)
\psline[linewidth=0.06,arrowsize=0.05291667 2.0,arrowlength=1.4,arrowinset=0.4]{cc->}(7.8,4.675)(8.0,2.875)
\psline[linewidth=0.06,arrowsize=0.05291667 2.0,arrowlength=1.4,arrowinset=0.4]{cc->}(10.2,2.275)(7.4,0.675)
\psline[linewidth=0.06,arrowsize=0.05291667 2.0,arrowlength=1.4,arrowinset=0.4]{cc->}(10.2,-1.525)(6.2,0.075)
\usefont{T1}{ptm}{m}{it}
\rput(3.3201563,5.18){singularity}
\usefont{T1}{ptm}{m}{it}
\rput(7.713594,5.18){last ray}
\usefont{T1}{ptm}{m}{it}
\rput(10.783906,2.58){dynamical horizon}
\usefont{T1}{ptm}{m}{it}
\rput(11.269531,-1.82){ingoing matter}
\psline[linewidth=0.08](0.0,0.275)(5.8,-5.325)
\psline[linewidth=0.08](5.8,-5.325)(11.2,0.475)
\psline[linewidth=0.08](11.2,0.475)(8.4,3.075)
\psline[linewidth=0.08](0.0,0.275)(2.8,3.075)
\end{pspicture} 
\caption{ The Penrose diagram of the CGHS space-time at the mean field approximation (MFA) level for an 
incoming $\delta-$function matter wave on null coordinates, showing the main features of the model. Unlike the 
$3+1$ dimensional case, here there are two past null infinities (left $\scri^-_L$ and right $\scri^-_R$) and two
future null infinities (left $\scri^+_L$ and right $\scri^+_R$), since the uncompact spatial coordinate is in the 
range $(-\infty,\infty)$, compared to a radial coordinate with range $[0,\infty)$ in $3+1$ dimensions.
The incoming $\delta-$function matter wave forms a black hole. 
To the left of the matter wave, the space-time is exactly flat, and in the vicinity of $\scri^-_R$ 
and $\scri^+_R$ the space-time is asymptotically flat. The singularity and the dynamical horizon (dashed line) 
meet at finite $z^{\pm}$. The last ray is the null line connecting this point to $\scri^+_R$.
Within the MFA only the region of spacetime to the  causal past of the last ray and singularity 
can (uniquely) be determined.} 
\label{cghs_sketch}
\end{figure}
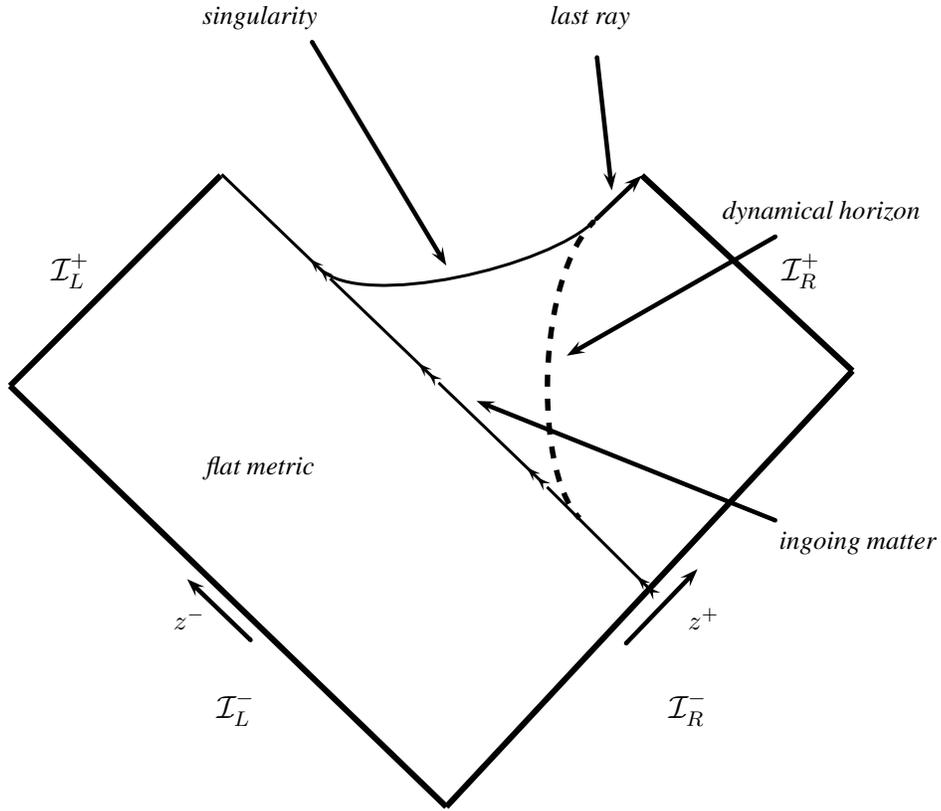

We are interested in metrics $g^{ab}$ that approach a Minkowski metric $\eta^{ab}$ at past null infinity. 
We will denote the null coordinates of $\eta$ as $z^{\pm}$ (see Fig.~\ref{cghs_sketch}).
Defining the fields $\Phi$ and $\Theta$ via
\be\label{defs}  \Phi \equiv e^{-2\phi}\quad {\rm and}\quad g^{ab} \equiv
\Theta^{-1} \Phi\, \eta^{ab} \equiv \Omega\, \eta^{ab}\, , \ee
we can write the equations of motion in terms of a set of {\em evolution
equations}
\ba \label{evolution} \Box_{(g)}\, f = 0 &\Leftrightarrow&
\Box_{(\eta)}f = 0\nonumber\\
\Dp\,\Dm\, \Phi + \k^2 \Theta &=& G\, T_{+-}\nonumber\\
\Phi \Dp\,\Dm \ln \Theta &=& -G\, T_{+-}, \ea
and {\em constraint equations}
\ba \label{constraint} -\Dp^2\, \Phi + \Dp\,\Phi \Dp\, \ln \Theta &=& G
T_{++}\nonumber\\
-\Dm^2\, \Phi + \Dm\,\Phi \Dm\, \ln \Theta &=& G T_{- -},\ea
where $\Box_{(g)}$ ($\Box_{(\eta)}$) is the wave operator
with respect to the metric $g_{ab}$ ($\eta_{ab}$),
$T_{ab}$ is the scalar field stress-energy tensor with components
denoted by $T_{++},...$, and we use
the notation $\partial_-\equiv\partial/\partial_{z^-}$, and similarly
for $\partial_+$.
Classically (and at the tree-level) $T_{ab}$ is trace-free, hence
$T_{+-}$ vanishes. However at the one-loop level $T_{+-}$ picks
up a non-zero value due to the trace anomaly, which if we now 
consider the superposition of $N$ identical massless scalar fields is
\ba G\, T_{+-} = \frac{NG\hbar}{24}\, \Dp\,\Dm\, \left( \ln \Phi -\ln \Theta \right) \, \ea

In a characteristic initial value problem, we specify initial data on
a pair of intersecting, null hypersurfaces $z^+(z^-)=z^+_0$
and $z^-(z^+)=z^-_0$, to the causal future of their
intersection point $(z^+_0,z^-_0)$ (see~\cite{Winicour:1998tz} for a review).
Thus one can see where the constraint equations~(\ref{constraint}) receive
their name: for example, if we specify the scalar field $f$ (hence $T_{++},\ T_{--}$) and 
metric field $\Theta$ on these surfaces as initial data, we
are {\em not} free to choose $\Phi$, which is then given by integrating (\ref{constraint}).
The constraint equations
are {\em propagated} by the evolution equations~(\ref{evolution}), namely, if the constraints
are satisfied on the initial hypersurfaces, solving for the fields to the causal future 
using (\ref{evolution}) guarantees the constraints are satisfied for all time.
This is exactly true at the analytical level, though in a numerical evolution
this property of the field equations will in general only be satisfied to 
within the truncation error of the discretization scheme.

Let $(z^+_0,z^-_0)=(-\infty,-\infty)$ be the initial data surface,
and denote the choice of initial ingoing and outgoing scalar field profiles by
\ba
f(z^+,z^-=z^-_0) &\equiv& f_+(z^+)\\
f(z^+=z^+_0,z^-) &\equiv& f_-(z^-),
\ea
where $f_+(z^+_0)=f_-(z^-_0)$.
Note that due to the conformally invariant nature of the wave operator
in 2D (\ref{evolution}), the solution for $f$ over the entire
spacetime is simply $f(z^+,z^-)=f_+(z^+)+f_-(z^-)-f(z^+_0,z^-_0)$.
With these initial conditions, and the condition that the metric
approaches Minkowski on left- and right-past null infinity, the solution to the constraints 
are~\cite{ATV}:
\ba \label{sol1} \Theta (z^\pm) &=& - \k^2 x^+\,x^-\nonumber\\
\Phi(z^\pm) &=& \Theta (z^\pm) - \f{G}{2}\textstyle{\int_0^{x^+}}
\dd\bar{x}^+\, \textstyle{\int_0^{\bar{x}^+}} \dd
\bar{\bar{x}}^+\, (\partial
f_+/\partial \bar{\bar{x}}^+)^2  \nonumber\\
&-& \f{G}{2} \textstyle{\int_{0}^{{x}^-}} \dd {\bar{x}}^-\,
\textstyle{\int_{0}^{\bar{x}^-}} \dd\bar{\bar{x}}^-\, (\partial
f_-/\partial \bar{\bar{x}}^-)^2\, . \ea
where the notation $F(z^\pm)$ denotes evaluation of the
corresponding field $F$ on the given initial hypersurface, and
\be \k x^+ = e^{\k z^+}, \quad \k x^- = -e^{-\k
z^-}\, .\ee

For a first study, we will exclusively consider the case
\be \frac12 (\partial f_+/\partial x^+ )^2 = M\ \delta(x^+ - x^+_0 ) \ , \ee
with $x^+_0 =1\ (z^+_0=0)$, and no incoming matter from the left ($f_-=0$). This choice reduces the problem to evolving the fields $\Phi$ and $\Theta$ according to (\ref{evolution}) with the asymptotic initial 
conditions
\ba \label{initial_cond} \Theta (z^{\pm}) &=& e^{\kappa(z^+-z^-)} \nonumber\\
\Phi(z^{\pm}) &=&  e^{\kappa(z^+-z^-)} - M (e^{\kappa z^+} -1 )\ , \ea
for $z^+>0$, $z^- \to -\infty$. Both fields are 
trivially given by $e^{\kappa(z^+-z^-)}$ for $z^+<0$. With these restrictions, any space-time is defined by the 
two quantities $M$ and $N$. $M$ is also the Bondi mass of this system as $z^- \to -\infty$. 

The classical solution ($\hbar=0$) to the future of the delta-pulse matter wave is exactly given by (\ref{initial_cond}), though now valid everywhere within this domain and not just
the initial data surface. This spacetime contains an event horizon relative to
right-null infinity, and a singularity inside (to the left of) the horizon. In the 
mean field approximation, the black hole evaporates, and the event horizon
is replaced by a dynamical apparent horizon. It is expected that the full quantum theory
will resolve the singularity, however in the MFA there is still a singularity inside
the dynamical horizon. When the evaporation has proceeded to the point where the
dynamical horizon meets the singularity (see Fig.~\ref{cghs_sketch}), it becomes naked, 
i.e. visible to observers at $\scri^+_R$. The MFA equations cannot be solved beyond
this Cauchy horizon, which we call the {\em last ray}. It should be possible
to mathematically extend the spacetime beyond the last ray, in particular as the
geometry does not appear to be singular here (except at the point the dynamical
horizon meets the last ray), though we will not explore those issues here.

In all our simulations we use $G=\hbar=\kappa=1$. The CGHS model gives the same physics if $N$ and $M$ are 
scaled by the same number, and $N/24$ gives a natural scale for the unit mass. For the results presented 
here, we always use $N=24$. Hence, by macroscopic mass, we mean $M \gg 1$, and by sub-Planck-scale mass, we mean $M \ll 1$.

\section{The Numerical Calculation}\label{sec_prep}

In this section we describe several novel aspects of our solution scheme that allows
us to uncover the physics of 2D black hole evaporation within the CGHS model.
This includes compactification of the coordinates (Sec.~\ref{sec_compact}),
regularization of the fields (Sec.~\ref{sec_regular}), the discretization
and solution strategy (Sec.~\ref{sec_disc}), and a Richardson extrapolation
algorithm to increase the order of convergence of the base, 2nd order accurate
scheme (Sec.~\ref{sec_rich}). We also discuss in Sec.~\ref{sec_sol_scri} some
difficulties in naively attempting to solve the discrete equations
near null-infinity, and how we extract desired properties of the
solution near $\scri^+_R$ in Sec.~\ref{sec_extract_scri}.

\subsection{Compactification of the Coordinates}\label{sec_compact}
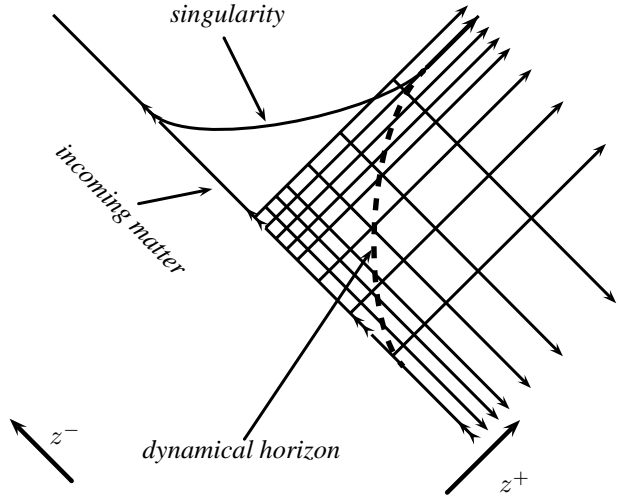
\begin{figure}
\psset{unit=1cm}
\begin{pspicture}(-2,-3.2773201)(8.114957,3.3032825)
\psline[linewidth=0.04,arrowsize=0.05291667 2.0,arrowlength=1.4,arrowinset=0.4]{-<<}(0.60938483,3.129961)(2.0199883,1.7121464)
\psline[linewidth=0.04,arrowsize=0.05291667 2.0,arrowlength=1.4,arrowinset=0.4]{-<<}(2.0199883,1.7121464)(3.4305916,0.29433185)
\psline[linewidth=0.04,arrowsize=0.05291667 2.0,arrowlength=1.4,arrowinset=0.4]{-<<}(3.4305916,0.29433185)(4.841195,-1.1234827)
\psline[linewidth=0.04,arrowsize=0.05291667 2.0,arrowlength=1.4,arrowinset=0.4]{-<<}(4.841195,-1.1234827)(6.251798,-2.5412972)
\psbezier[linewidth=0.072,linestyle=dashed,dash=0.16 0.16]{cc-}(5.5573134,2.4102368)(4.730255,1.5873849)(4.685286,-0.9667769)(5.264376,-1.548827)
\psbezier[linewidth=0.04]{cc-}(1.8789278,1.8539279)(2.443169,1.286802)(4.9496784,1.8056927)(5.5573134,2.4102368)
\psline[linewidth=0.06,arrowsize=0.05291667 2.0,arrowlength=1.4,arrowinset=0.4]{cc->}(5.5573134,2.4102368)(6.2662206,3.1155386)
\psline[linewidth=0.06,arrowsize=0.05291667 2.0,arrowlength=1.4,arrowinset=0.4]{cc->}(5.8257327,-3.2473202)(6.818203,-2.2598977)
\psline[linewidth=0.06,arrowsize=0.05291667 2.0,arrowlength=1.4,arrowinset=0.4]{cc->}(0.876362,-3.0932796)(0.03,-2.242591)
\psline[linewidth=0.04]{cc-}(3.4305916,0.29433185)(6.2662206,3.1155386)
\psline[linewidth=0.04,arrowsize=0.05291667 2.0,arrowlength=1.4,arrowinset=0.4]{cc->}(3.571652,0.1525504)(6.407281,2.9737573)
\psline[linewidth=0.04,arrowsize=0.05291667 2.0,arrowlength=1.4,arrowinset=0.4]{cc->}(3.7127123,0.01076895)(6.5483413,2.8319757)
\psline[linewidth=0.04,arrowsize=0.05291667 2.0,arrowlength=1.4,arrowinset=0.4]{cc->}(4.1358933,-0.41457543)(6.9715223,2.4066312)
\psline[linewidth=0.04,arrowsize=0.05291667 2.0,arrowlength=1.4,arrowinset=0.4]{cc->}(3.8537726,-0.1310125)(6.6894016,2.6901941)
\psline[linewidth=0.04,arrowsize=0.05291667 2.0,arrowlength=1.4,arrowinset=0.4]{cc->}(4.5590744,-0.8399198)(7.3947034,1.981287)
\psline[linewidth=0.04,arrowsize=0.05291667 2.0,arrowlength=1.4,arrowinset=0.4]{cc->}(5.123316,-1.4070456)(7.958945,1.4141611)
\psline[linewidth=0.04,arrowsize=0.05291667 2.0,arrowlength=1.4,arrowinset=0.4]{cc->}(3.4313128,0.57717365)(6.39358,-2.4002368)
\psline[linewidth=0.04,arrowsize=0.05291667 2.0,arrowlength=1.4,arrowinset=0.4]{cc->}(3.5730941,0.718234)(6.5353613,-2.2591765)
\psline[linewidth=0.04,arrowsize=0.05291667 2.0,arrowlength=1.4,arrowinset=0.4]{cc->}(3.7148757,0.85929435)(6.6771426,-2.1181161)
\psline[linewidth=0.04,arrowsize=0.05291667 2.0,arrowlength=1.4,arrowinset=0.4]{cc->}(3.9984386,1.141415)(6.9607058,-1.8359956)
\psline[linewidth=0.04,arrowsize=0.05291667 2.0,arrowlength=1.4,arrowinset=0.4]{cc->}(4.423783,1.564596)(7.3860497,-1.4128146)
\psline[linewidth=0.04,arrowsize=0.05291667 2.0,arrowlength=1.4,arrowinset=0.4]{cc->}(5.13269,2.2698977)(8.094957,-0.70751286)
\psline[linewidth=0.04,arrowsize=0.05291667 2.0,arrowlength=1.4,arrowinset=0.4]{cc->}(3.2895312,0.43611333)(6.12516,3.2573202)
\psline[linewidth=0.04,arrowsize=0.05291667 2.0,arrowlength=1.4,arrowinset=0.4]{cc->}(2.871398,2.8413503)(3.4341972,1.7085408)
\psline[linewidth=0.04,arrowsize=0.05291667 2.0,arrowlength=1.4,arrowinset=0.4]{cc->}(3.1405387,-2.533365)(4.8440795,0.007884475)
\usefont{T1}{ptm}{m}{it}
\rput(2.9260743,3.1182823){singularity}
\usefont{T1}{ptm}{m}{it}
\rput(6.7468553,-3.0817175){$z^+$}
\usefont{T1}{ptm}{m}{it}
\rput(0.7668555,-2.4417176){$z^-$}
\usefont{T1}{ptm}{m}{it}
\rput(3.1093554,-2.6817176){dynamical horizon}
\usefont{T1}{ptm}{m}{it}
\rput{-45.660347}(0.092023715,1.2173457){\rput(1.4812305,0.5182824){incoming matter}}
\psline[linewidth=0.04,arrowsize=0.05291667 2.0,arrowlength=1.4,arrowinset=0.4]{cc->}(1.7254492,0.6132824)(2.7254493,0.8132824)
\end{pspicture} 
\caption{A schematic view of the positions of the grid lines on the uncompactified space. Lines are 
concentrated near the last ray, where we need higher resolution. They become distant as one approaches the null 
infinities.} 
\label{cghs_mesh_uncompact}
\end{figure}
Rather than discretizing the equations with respect to the $z^+,z^-$ coordinates,
we introduce a compactified coordinate system $z^+_c \in [0, \frac12]$ and $z^-_c \in [0, 1]$. 
Use of compact coordinates is important for a couple of reasons, and essential for the $M \gg 1$ case. 
First, to understand the asymptotic structure of the spacetime approaching $\scri_R^+$, it is useful
to have the computational domain include $\scri_R^+$.
Second, the uncompactified coordinate $z^-$ 
is adapted to the flat metric near $\scri^-_L$; however, it turns out that most of the interesting features of 
black hole evaporation near the dynamical horizon occur in an exponentially small 
region $\Delta z^- \sim \kappa^{-1} e^{-GM/\kappa}$ before 
the last ray. One can think of this as essentially due to gravitational redshift. 
Classically (without evaporation), the redshift causes arbitrarily small lengths scales
near the horizon to be expanded to large scales near $\scri^+_R$.
Naively one might have expected that evaporation changes this pictures
completely (as suggested by the Penrose diagram in Fig.~\ref{cghs_sketch}). Instead, what we
find is that although there is not an arbitrarily large redshift once back-reaction
is included, there is still an exponential growth of scales, with the growth rate
proportional to the mass of the black hole as indicated above.

Thus, a uniform discretization in $z^-$ that is able to resolve both the early
dynamics near $\scri_R^-$, yet can adequately uncover the exponentially small scales (as measured
in $z^-$) of the late-time evaporation, will (for large $M$) result in a mesh too large to be able to solve 
the equations on using contemporary computer systems.
To overcome this problem, we introduce a non-uniform compactification in $z^-$, schematically
illustrated in Fig.~\ref{cghs_mesh_uncompact}), that provides sufficient resolution
to resolve the spacetime near the last ray, yet does not over-resolve the region approaching
$\scri_R^-$. Specifically, the transformation from $z^-$ to $z^-_c$ we use is as follows.
First, we relate the uncompactified $z^-$ to an auxiliary (non-compact) coordinate $\bar{z}^-$ by
\begin{equation} 
z^- = \bar{z}^-\ \left(\frac{\bar{z}^- - L_R^{-1/2} }{\bar{z}^- - L_R^{1/2}}\right) +z^-_{s,est} 
\end{equation}
where $z^- \in (-\infty,z^-_{s,est}]$ and $\bar{z}^- \in (-\infty,0]$. $z^-_{s,est}$ is an estimate of the 
$z^-$ coordinate of the last ray. This is also the earliest time in $z^-$ that we
will encounter the spacetime singularity, and at present we do not continue the computation past 
this point (the compactification functions can readily be adjusted to cover $z^- 
\in (-\infty , \infty)$ ). This way, the region near the last ray ($z^- \approx z^-_{s,est}$, $\bar{z}^- 
\approx 0$) is resolved by a factor of $L_R$ more than the regions away from the last ray. Next, we 
convert the auxiliary $\bar{z}^-$ to a compact coordinate $z_c^-$

\begin{equation}
\bar{z}^- = -e^{-S \tan(\pi z_c^- -\pi/2)}+ L_c (z_c^--1), \ .
\end{equation} 
where $S$ and $L_c$ are constants.
This way, the last ray is located near $z_c^-=1$. The relation between $\bar{z}^-$ and $z_c^-$ is forced to be linear 
near the last ray through the $L_c$ term. Our grid has a fixed step size $\Delta z_c^- = h$ in the compactified 
coordinate $z_c^-$, which corresponds to $\Delta z^- = L_c/L_R \ h$ in uncompactified coordinates near $z_c^-=1$.

For the highest mass macroscopic black hole discussed here, $M=16$, we set $L_R=10^9$, while for the lowest
mass of $M=2^{-10}$, we use $L_R=10^2$. We use $L_c=4.096 \times 10^{-9}$, which can be adjusted together 
with $L_R$ to obtain the desired resolution near the last ray. Note that $\Delta z^- \approx 10^{-18} h$ for the highest mass case; 
such a disparity in scales would have been difficult to achieve if we had used $z^-$ as our coordinate even {\em with}
a standard adaptive mesh refinement algorithm.
We choose $S$ to be between $1$ and $5$, the particular value of which is not too essential.

In the $+$ direction, for $M\tilde{>} 1$, we compactify using
\be z^+ = M \tan(\pi z^+_c) \hspace{1cm} M\tilde{>} 1 \ , \ee 
with the factor of $M$ ensuring that the singularity is not too close to the $\scri^+_R$ edge of the mesh. For $M \ll 1$, the 
singularity appears very close to $z^+=0$, so to resolve this region, we employ
\be z^+ = C_{z^+_c} \tan^p(\pi z^+_c) \hspace{1cm} M \ll 1 \ , \ee
where $C_{z^+_c}$ and $p$ are appropriate constants that again keep the singularity near the middle of the range of $z^+_c$. For 
$M=2^{-10}$, we use $C_{z^+_c}=\frac{1}{7000}$ and $p=7$.

\subsection{Regularization of the Fields}\label{sec_regular}

It is clear from (\ref{initial_cond}) that the fields diverge exponentially at $\scri^-_R$ and analytical results show that they 
also diverge at $\spr$ \cite{ATV}. For a numerical solution then, we defined regularized field variables which are finite 
everywhere 
\ba 
\Phi &=& e^{\kappa(z^+-z^-)}\ (1+\bar{\phi}) - M (e^{\kappa z^+} -1 ) \nonumber\\
&=& e^{\kappa(z^+-z^-)}\ (1+\bar{\phi} +\bar{\phi}_0) \nonumber\\
\Theta &=& e^{\kappa(z^+-z^-)}\ (1+\bar{\theta}) \ ,
\ea
with $\bar{\phi}_0 = -M\ e^{\kappa z^-}(1 - e^{- \kappa z^+} )$. Aside from removing the divergent component $e^{\kappa(z^+-z^-)}$, 
this definition also removes the exact classical solution $M (e^{\kappa z^+} -1 )$ from $\Phi$. The reason
for doing this came from preliminary studies that showed
deviations in $\Phi$ were small compared to the classical metric for macroscopic black holes. In terms of the new variables, 
equations (\ref{evolution}) read	
\ba
(1+\bar{\theta})^2(1+\bar{\phi}+\bar{\phi}_0)^2 \hspace{4.5cm} && \nonumber \\
\times \left[ \partial_+ \partial_- \bar{\phi}
		- \kappa \partial_+ \bar{\phi}
		+ \kappa \partial_- \bar{\phi}
		-\kappa^2 \bar{\phi}
		+\kappa^2 \bar{\theta} \right] -Q(\bar{\phi}, \bar{\theta})
		&=& 0 \nonumber \\
\label{eq:evolution_regularized1}
\ea
and
\ba		
(1+\bar{\phi}+\bar{\phi}_0)^3 \big[(1+\bar{\theta}) \partial_+ \partial_- \bar{\theta} -\partial_+ \bar{\theta} \partial_- \bar{\theta}  \big]
+Q(\bar{\phi}, \bar{\theta})	&=& 0 \nonumber \\
\label{eq:evolution_regularized2}				
\ea
with
\ba
Q(\bar{\phi}, \bar{\theta}) &=& \frac{NG\hbar}{24} e^{\kappa (z^- -z^+)} \nonumber \\
\times \bigg\{ && (1+\bar{\theta})^2 \big[(1+\bar{\phi}+\bar{\phi}_0)\ \partial_+ \partial_- ( \bar{\phi}+\bar{\phi}_0)  \big] \nonumber \\
	&-& (1+\bar{\theta})^2 \big[\partial_+ (\bar{\phi}+\bar{\phi}_0)\ \partial_- (\bar{\phi}+\bar{\phi}_0) \big] \\
	&-&(1+\bar{\phi}+\bar{\phi}_0)^2 \big[(1+\bar{\theta}) \partial_+ \partial_- \bar{\theta} -\partial_+ \bar{\theta} \partial_- \bar{\theta}  \big] \bigg\} \nonumber \label{disc_eqns}
\ea

\subsection{Discretization and Algebraic Manipulation}\label{sec_disc}
\begin{figure}
\psset{unit=0.85cm}
\begin{pspicture}(0,-5.615418)(13.979246,5.615418)
\definecolor{color2780}{rgb}{0.48627450980392156,0.4666666666666667,0.4666666666666667}
\definecolor{color2803}{rgb}{0.5294117647058824,0.5176470588235295,0.5176470588235295}
\rput{45.923378}(1.7130033,-4.043097){\psframe[linewidth=0.06,dimen=outer](9.567839,3.98)(1.6478391,-3.94)}
\psline[linewidth=0.1cm,linecolor=color2780,arrowsize=0.05291667cm 2.0,arrowlength=1.4,arrowinset=0.4]{-<<}(2.7541983,2.7824788)(4.1910186,1.3912394)
\psbezier[linewidth=0.072,linestyle=dashed,dash=0.16cm 0.16cm]{cc-}(7.7192564,1.8724246)(6.961804,1.0901557)(7.049535,-1.3765948)(7.6393876,-1.9477352)
\psbezier[linewidth=0.072]{cc-}(4.1910186,1.3912394)(4.7657466,0.8347437)(7.1627607,1.2976965)(7.7192564,1.8724246)
\psline[linewidth=0.06cm,arrowsize=0.05291667cm 2.0,arrowlength=1.4,arrowinset=0.4]{cc->}(7.7192564,1.8724246)(8.554,2.7345169)
\psline[linewidth=0.06cm,arrowsize=0.05291667cm 2.0,arrowlength=1.4,arrowinset=0.4]{cc->}(7.6849685,-4.775795)(8.658836,-3.7700207)
\psline[linewidth=0.06cm,arrowsize=0.05291667cm 2.0,arrowlength=1.4,arrowinset=0.4]{cc->}(3.1372824,-3.4346945)(2.27519,-2.599951)
\psline[linewidth=0.04cm]{cc-}(6.346249,-0.6956197)(9.128728,2.178021)
\psline[linewidth=0.04cm]{cc-}(7.0646596,-1.3912394)(9.847138,1.4824014)
\psline[linewidth=0.04cm]{cc-}(7.7830696,-2.0868592)(10.565549,0.7867816)
\psline[linewidth=0.04cm]{cc-}(9.1971,-2.0640688)(6.3234587,0.7184102)
\psline[linewidth=0.04cm]{cc-}(9.892719,-1.3456585)(7.0190783,1.4368204)
\psline[linewidth=0.04cm]{cc-}(10.588339,-0.6272483)(7.7146983,2.1552305)
\psline[linewidth=0.04cm]{cc-}(5.627839,0.0)(8.410318,2.8736408)
\rput{45.923378}(1.7038282,-9.333409){\psframe[linewidth=0.06,dimen=outer](13.369401,-1.1530341)(10.3635025,-4.158933)}
\psline[linewidth=0.04cm]{cc-}(7.7374887,0.7412007)(9.735892,-2.6152706)
\psline[linewidth=0.04cm]{cc-}(8.433108,1.4596108)(11.935892,-0.6152706)
\psline[linewidth=0.1cm,linecolor=color2780,arrowsize=0.05291667cm 2.0,arrowlength=1.4,arrowinset=0.4]{-<<}(4.1910186,1.3912394)(5.627839,0.0)
\psline[linewidth=0.1cm,linecolor=color2780,arrowsize=0.05291667cm 2.0,arrowlength=1.4,arrowinset=0.4]{-<<}(5.627839,0.0)(7.0646596,-1.3912394)
\psline[linewidth=0.1cm,linecolor=color2780,arrowsize=0.05291667cm 2.0,arrowlength=1.4,arrowinset=0.4]{-<<}(7.0646596,-1.3912394)(8.50148,-2.7824788)
\psline[linewidth=0.1cm,linecolor=color2803]{cc-}(8.50148,-2.7824788)(11.283958,0.0911619)
\psline[linewidth=0.04cm,arrowsize=0.05291667cm 2.0,arrowlength=1.4,arrowinset=0.4]{cc->}(3.9358923,-0.8152706)(5.7358923,-0.2152706)
\psline[linewidth=0.04cm,arrowsize=0.05291667cm 2.0,arrowlength=1.4,arrowinset=0.4]{cc->}(3.9358923,-0.8152706)(9.5358925,-1.6152706)
\usefont{T1}{ptm}{m}{it}
\rput(1.8772986,-3.3102705){$z_c^- , i$}
\usefont{T1}{ptm}{m}{it}
\rput(8.927299,-4.5102706){$z_c^+ , j$}
\usefont{T1}{ptm}{m}{it}
\rput(3.0033925,-0.5102706){initial conditions}
\usefont{T1}{ptm}{m}{it}
\rput(5.2841735,-2.7102706){flat metric}
\usefont{T1}{ptm}{m}{it}
\rput(11.187299,-1.7102706){\large $h$}
\usefont{T1}{ptm}{m}{it}
\rput(10.987299,-3.5102706){\large $h$}
\usefont{T1}{ptm}{m}{it}
\rput(12.387299,-1.7102706){\large $h$}
\usefont{T1}{ptm}{m}{it}
\rput(12.587298,-3.5102706){\large $h$}
\psline[linewidth=0.04cm,arrowsize=0.05291667cm 2.0,arrowlength=1.4,arrowinset=0.4]{cc->}(5.335892,2.3847294)(5.5358925,1.1847295)
\psline[linewidth=0.04cm,arrowsize=0.05291667cm 2.0,arrowlength=1.4,arrowinset=0.4]{cc->}(9.5358925,3.1847293)(7.1358924,0.5847294)
\usefont{T1}{ptm}{m}{it}
\rput(5.1365175,2.8897295){singularity}
\usefont{T1}{ptm}{m}{it}
\rput(9.853393,3.4897294){dynamical horizon}
\usefont{T1}{ptm}{m}{it}
\rput(11.837298,-4.9102707){\footnotesize $i-1,j-1$}
\usefont{T1}{ptm}{m}{it}
\rput(9.487299,-3.1102705){\footnotesize $i,j-1$}
\usefont{T1}{ptm}{m}{it}
\rput(12.337298,-0.3102706){\footnotesize $i,j$}
\usefont{T1}{ptm}{m}{it}
\rput(13.087298,-2.7102706){\footnotesize $i-1,j$}
\end{pspicture}
\caption{ The grid structure for the numerical calculation. We use a fixed-step-size mesh based on the 
compactified coordinates $z_c^{\pm}$, where the step sizes in both directions are equal. The emphasis on the regions 
where the fields rapidly change is attained using the compactification of the coordinates (see Fig.
\ref{cghs_mesh_uncompact}). The flat region before the matter pulse and the region beyond the last ray are not covered by the mesh.} 
\label{cghs_mesh}
\end{figure}
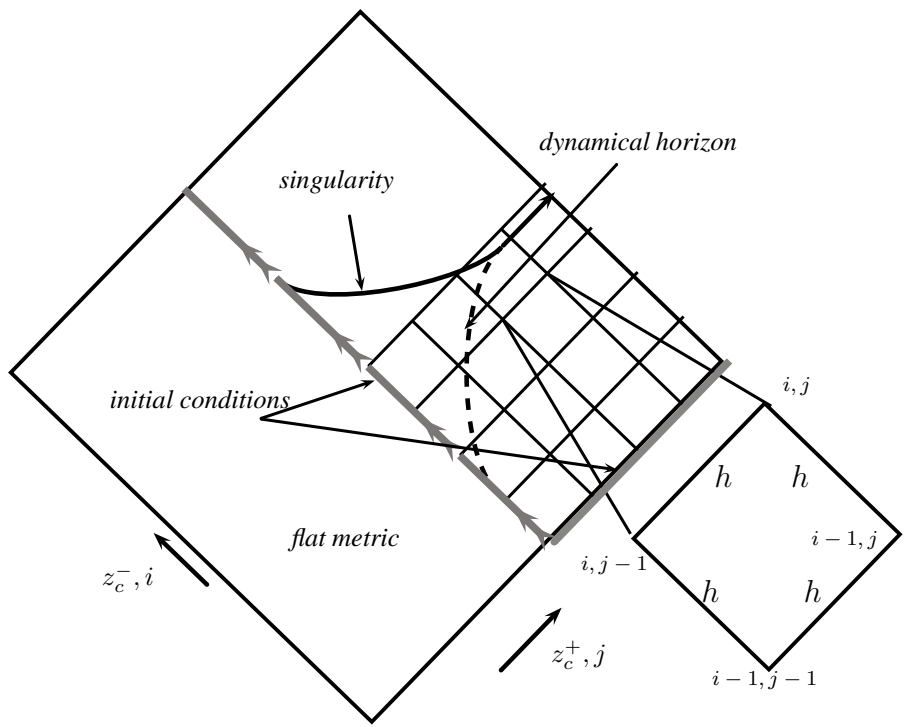

We discretize the compactified coordinate domain as depicted in Fig.~\ref{cghs_mesh}. 
A field $\alpha(z^+_c,z^-_c)$ is represented by a discrete mesh of values 
$\alpha_{i,j}$, where the indices $i,j$ are integers, and related to the null 
coordinates through
\ba z^-_c &= i h \hspace{1cm} 0 \leq i \leq n_{p} \nonumber \\ 
 z^+_c &= j h \hspace{1cm} 0 \leq j \leq \frac{n_{p}}{2} \ ,
\ea
where $h=n_{p}^{-1}$ is the step size in both of the compactified null coordinates.
In order to solve the evolution equations numerically, we convert the differential equations to
difference equations by using standard, second order accurate ($\mathcal{O}(h^2)$), centered stencils:
\ba 
 \alpha \big|_{i-\frac{1}{2},j-\frac{1}{2}} &\approx& \frac{\alpha_{i,j} +\alpha_{i-1,j}+\alpha_{i,j-1}+\alpha_{i-1,j-1}}{4} \nonumber \\
\partial'_+ \alpha \big|_{i-\frac{1}{2},j-\frac{1}{2}} &\approx& \frac{\alpha_{i,j} +\alpha_{i-1,j}-\alpha_{i,j-1}-\alpha_{i-1,j-1}}{2h} \nonumber \\
\partial'_- \alpha \big|_{i-\frac{1}{2},j-\frac{1}{2}} &\approx& \frac{\alpha_{i,j} -\alpha_{i-1,j}+\alpha_{i,j-1}-\alpha_{i-1,j-1}}{2h} \nonumber \\
\partial'_+\partial'_- \alpha \big|_{i-\frac{1}{2},j-\frac{1}{2}} &\approx& \frac{\alpha_{i,j} -\alpha_{i-1,j}-\alpha_{i,j-1}+\alpha_{i-1,j-1}}{h^2}, \nonumber \\
\label{eq:difference_equations}
\ea
where we have introduced the notation
\be
\partial'_{\pm} \equiv \frac{\partial}{\partial z^{\pm}_c} = \frac{\partial z^{\pm}}{\partial z^{\pm}_c} \frac{\partial}{\partial z^{\pm}}
= \frac{\partial z^{\pm}}{\partial z^{\pm}_c} \partial_{\pm} \ .
\ee
Once discretized, (\ref{eq:evolution_regularized1}) and (\ref{eq:evolution_regularized2}) give two polynomial 
equations which can be numerically solved for $\bar{\theta}_{i,j}$ and $\bar{\phi}_{i,j}$, if the field values are 
known at the grid points $(i,j-1), (i-1,j), (i-1,j-1)$. This way, knowing the boundary conditions at $z^+=0\ (j=0)$ 
and $z^-=-\infty \ (i=0)$, we can calculate the field values at all points of the grid one by one, starting at $(1,
1)$. 

Instead of solving for the two variables simultaneously (e.g. using a two dimensional Newton's method), we sum the 
equations (\ref{eq:evolution_regularized1}, \ref{eq:evolution_regularized2}), which allows us to explicitly express 
$\bar{\phi}_{i,j}$ in terms of a rational function of $\bar{\theta}_{i,j}$. We then insert this expression for 
$\bar{\phi}_{i,j}$ into (\ref{eq:evolution_regularized1}){\footnote{alternatively, any other independent linear 
combination of the equations can be used}. This way, we obtain a single variable, $10^{th}$ order polynomial equation for
$\bar{\theta}_{i,j}$. We solve this equation numerically using 
Newton's method, and then calculate  $\bar{\phi}_{i,j}$ directly using the aforementioned rational function. Many 
other techniques are available for finding the roots of polynomials in one variable. For instance, we also implemented
Laguerre's method, which gave similar results in terms of robustness and computation time.

\subsection{Richardson extrapolation with intermittent error removal}\label{sec_rich}
For any function $\alpha$ numerically calculated on a null mesh of step size $h$ in both directions, and
with central differences as in (\ref{eq:difference_equations}), we have a Richardson expansion
\be\label{rich}
\alpha_h = \alpha + c_2 h^2 + c_4 h^4 + c_6 h^6 + \mathcal{O}(h^8)
\ee 
where $\alpha$ is the exact solution, $\alpha_h$ is the numerically obtained solution and $c_i$ are {\em error functions}.
$\alpha$, $\alpha_h$, and $c_i$ are all functions of $z^{\pm}$ (we omit the explicit dependence for clarity), and $\alpha,c_i$ are 
independent of $h$. Note that we cannot {\em prove} such an expansion exists for the class of non-linear
equations we are solving, in particular if no assumptions on the smoothness of the initial data are made. Furthermore,
we know the solutions generically develop singularities, thus the above series can only have a limited radius
of convergence for generic initial data. Nevertheless, we will {\em assume} the expansion exists, and then,
via convergence tests, check whether the solutions we obtain are consistent with the expansion.

The use of second order finite difference stencils is responsible for the leading order quadratic convergence of the above expansion.
However, using numerical solutions obtained on meshes with different discretization scales, one can obtain higher order convergence
by using the well known Richardson extrapolation. For example, a fourth order convergent solution 
$\alpha_{h,h/2}$ can be obtained from the following superposition of two approximate second order convergent
solutions $\alpha_{h/2}$ and $\alpha_{h}$ : $\alpha_{h,h/2} = (4\alpha_{h/2} - \alpha_{h})/3 = \alpha + \mathcal{O}(h^4)$.
In theory (for sufficiently smooth solutions), $2n$-th order convergence can be obtained by an
appropriate superposition of $n$ second order accurate solutions, each obtained with a different mesh spacing.
As we describe in more detail below, we use four successively finer meshes to obtain solutions that converge to
$\mathcal{O}(h^8)$ on the points of the coarsest mesh.

Fields in the CGHS model present singular behavior, and since the position where the singularity first
appears is a (convergent) function of the mesh size, the method of superposing solutions of different meshes breaks down 
at the first time the singularity appears on {\em any} of the superposed meshes. Typically, the singularity 
first appears on the coarsest mesh, and thus our domain of integration is fundamentally restricted by 
our proximity to the singularity on the coarsest mesh.  
Many of the physical phenomena we are interested in occur in this region, thus a direct use of Richardson 
extrapolation for solutions over the entire computational domain does not significantly improve our results. 
To circumvent this problem, as described in more detail in the next few paragraphs, we instead break up the computational
domain into a series of short strips in $z^-_c$. In each strip we evolve 4 meshes, apply Richardson extrapolation
to the solution obtained at the end of the evolution, then use this corrected solution as initial data for
all four meshes on the next, adjacent strip. In this way the mismatch in the location of the singularity
amongst the four resolutions is confined to be less than the size of the strip, which we can
adjust as needed.

Our Richardson extrapolation algorithm proceeds as follows. We divided the entire grid 
into $L$ equal regions along $z^-_c$ such that grid points $i$ along the corresponding direction
with $\frac{l}{L} n_{p} \leq i \leq \frac{l+1}{L} n_{p}$, $0 \leq l \leq L$ comprise the $l^{th}$ region. 
Note that regions coincide at the boundaries, and here indices $i$ and the total number
of points $n_{p}$ are relative to the coarsest mesh---for finer meshes these numbers
should be scaled as appropriate so that the $l^{th}$ strip spans the {\em same coordinate volume} 
for each resolution. In the $l^{th}$ region  
\begin{enumerate}
\item We evolve the fields independently on four successively finer meshes of step size $h, h/2, h/4$ and $h/8$, 
and stop the evolution at the end of the region ($i = \frac{l+1}{L} n_{p}$).

\item At points coincident with the coarsest resolution, we calculate the appropriate superposition of the four meshes 
to give $\mathcal{O}(h^8)$ accurate values of the 
fields ($\bar{\phi}$ and $\bar{\theta}$), and store these values on the coarsest mesh as our result.

\item On the last ($i = \frac{l+1}{L} n_{p}$) line of the region $l$, we also calculate the functions 
$c_k(z^{\pm})$ to accuracy $\mathcal{O}(h^{8-k})$ on the coarsest mesh. We then interpolate the functions $c_k$ to the three finer 
meshes using successive degree four Lagrange interpolating polynomials. Using these interpolated $c_k$ values, we 
correct the field values on the finer meshes using (\ref{rich}).  A Lagrange polynomial of degree $4$ introduces
an error of order $\mathcal{O}(h^5)$, and hence through the $c_2$ term an error of $\mathcal{O}(h^7)$ will
be introduced into the finer mesh solutions. A higher order interpolating polynomial could reduce the
error, though we found that a global $\mathcal{O}(h^7)$ scheme is sufficient for our purposes.

\item We use the $\mathcal{O}(h^7)$ accurate field values on the last ($i = \frac{l+1}{L} n_{p}$) line as the 
initial data for the next ($(l+1)^{th}$) region, and repeat the procedure for this region starting from Step $1$.
\end{enumerate}
By updating the fields to more accurate values at the end of each region, the accuracy of the position of the singularity 
in the coarsest mesh is improved significantly, and the problem of the breakdown of the superposition near the singularity 
is overcome.  

We are not aware of any studies on the theoretical stability and accuracy of this modified Richardson extrapolation method, 
though our convergence and independent residual analysis, described in Sec.~\ref{sec_tests}, shows that it works quite well,
giving (for the most part) the expected order of convergence.

Implementing this method, we are able to reduce the truncation error down to the level of round-off error
using ``modest'' resources on a single, desktop style CPU. More precisely, we used $80$-bit 
long double precision, which theoretically has a round-off error at the level of $\sim 10^{-19}$, however
our Newton iteration only converged if we set the 
accuracy of the iteration to $\sim 10^{-16}$, which was the ultimate source of the error in the calculated regularized 
field values. When we say ``round-off error'' then, we will mean this latter value rather than the 
value of $\sim 10^{-19}$ one might expect from $80$-bit precision.

\subsection{Evolution near $\scri$}\label{sec_sol_scri}
The boundary conditions on $z^+=0$ are those of the vacuum and translate to
\be \bar{\phi}(z^+=0, z^-) = \bar{\theta}(z^+=0, z^-) = 0 \ . \ee 
For the $\scri_R^-$ boundary, (\ref{initial_cond}) translates to 
\be \bar{\phi}(z^+, z^- = -\infty) = \bar{\theta}(z^+, z^-=-\infty) = 0 \ . \ee 
The $e^{\kappa z^-}$ factors in the evolution equations (\ref{disc_eqns}) are interpreted as $0$ if 
$e^{\kappa z^-}$ is less than the smallest magnitude floating point number allowed by machine precision, 
which occurs for $z^- < z^-_{prec}$ for some $z^-_{prec}$, thus, in the region $z^- < z^-_{prec}$,
the evolution equations are trivially solved by the initial conditions $\bar{\theta}=\bar{\phi}=0$.
This means it would make no difference if we imposed the $\scri_R^-$ boundary 
conditions on some other constant $z^- < z^-_{prec}$ line. Moreover, even if we impose the $\scri^-_R$ 
boundary conditions on a constant $z^-$ line with $z^-> z^-_{prec}$, the error introduced is exponentially 
small~\cite{PiranStrominger} and negligible compared to the truncation error for a certain range of $|z^-|$. Our numerical method described in Sec.~\ref{sec_disc} sometimes fails to produce a solution for the fields in the 
early stages of the evolution near $\scri_R^+$ if we begin the evolution in the region $z^- < z^-_{prec}$. We surmise the failure occurs near the line $z^- = z^-_{prec}$. In such cases of failure, we begin the evolution at $z^- \sim - 5 \times 10^3$, which introduces a 
completely negligible error.

A related problem is that Newton's method also sometimes cannot converge to a solution
for $\bar{\theta}$ near $\scri_R^+$, even well before the last ray. Nevertheless, we were able to evolve 
the fields sufficiently close to $\scri_R^+$ to extract all the important asymptotic behavior, as 
described in the next sub-section.

\subsection{Asymptotic Behavior}\label{sec_extract_scri}
Many of the physical quantities of black hole evaporation are related to the asymptotic behavior of $\Phi$ and $\Theta$ near $\scri^+_R$. The dynamical fields admit the asymptotic expansions \cite{ATV}
\ba \Phi &=& A(z^-) e^{\k z^+} + B(z^-) +
\mathcal{O}(e^{-\k z^+})\nonumber\\
\Theta &=& \ub{A}(z^-) e^{\k z^+} + \ub{B}(z^-) + \mathcal{O}(e^{-\k
z^+})\, .\ea
which are used to calculate the affine parameter $y^-$ on $\scri^+_R$ through $\kappa \exp(-\kappa y^-)=A(z^-)$. 
The equations also admit a balance law on $\scri_R^+$ 
\ba \label{balance}\f{d}{\dd\y^-}\big[ \f{\dd{B}}{\dd\y^-} + \k
{B}\, +\, \f{N \hbar G}{24}\, \big(\f{\dd^2\y^-}{\dd z^{-2}}\,
(\f{\dd\y^-}{\dd z^-})^{-2}\,
\big)\,\big]
= - \f{N \hbar G}{48}\, \big[\f{\dd^2\y^-}{\dd z^{-2}}\,
(\f{\dd\y^-}{\dd z^-})^{-2}\,\,\big]^2 \ea
We identify the expression in square brackets to the left of the equal sign as $GM_B$, with $M_B$ identified as the 
Bondi mass as in \cite{ATV}. We will call the term on the right hand side the Ashtekar-Taveras-Varadarajan flux, 
$F_{ATV}$. Asymptotic coefficients are related to the regularized fields through
\ba\label{asymp_eqns}
A(z^-) &=& e^{-\kappa z^-} \left( 1+\bar{\phi}(z^+=\infty, z^-)\right) -M \\
B(z^-) &=& \lim_{z^+ \to \infty} e^{\kappa z^+} \left( \bar{\phi}(z^+=\infty, z^-) - \bar{\phi}(z^+, z^-) \right) + M \nonumber
\ea
 
As mentioned in the previous section, we are not able to calculate the fields exactly on $\scri_R^+$, and
evaluating the above on a line of constant $z^+$ will introduce an error of the order $e^{-\kappa z^+}$.
However, it is adequate to evaluate the above at sufficiently large $z^+$ such that this error is less
than the truncation error. It turns out the Newton iteration only breaks down well into the region
where the truncation error dominates, and we calculate $A$ on a line of constant $z^+$ in this region.

To calculate the limit in $B$ numerically, we need (at least) two values of $z^+$ for each value of $z^-$. From an analytical point of view, it is most desirable to use two $z^+$ values as large as possible. However, $B$ is expressed as 
the asymptotically diverging factor $e^{\kappa z^+}$ multiplied by an asymptotically vanishing one, 
and calculating this via finite precision numerics could introduce a large round-off error due to catastrophic cancellation.
We thus evaluate $B$ using two $z^+ = const$ lines, one the line we used to calculate $A$, the other chosen such that $\kappa z^+$ is large enough that the fields are in the asymptotic region but it is also sufficiently away from the other $z^+ = const$ line that catastrophic cancellation is not a major issue. Particular values of $z^+$ are not important.
To calculate $M_B$ and the ATV flux (\ref{balance}), we use nine-point stencils to calculate the first and second derivatives 
with respect to $z^-_c$ (applying the chain rule to obtain derivatives with respect
to $z^-$), which have an accuracy of $\mathcal{O}(h^8)$, keeping the theoretical accuracy 
of our numerical integration scheme.

Note that since $B$ is sub-leading relative to $A$ in the asymptotic
expansion (\ref{asymp_eqns}), $A$, hence $y^-$, can be calculated
more accurately. Thus, in practice we calculate the Bondi
mass $M_B$ by numerically integrating the ATV flux rather than directly
evaluating the left hand side of (\ref{balance}). 

\section{Numerical Tests}\label{sec_tests}

In this section we present a few sample solutions to the CGHS model
in the mean field approximation, and results from an array of
tests we performed to ensure we are solving the equations correctly.

\subsection{Sample evolutions}\label{sec_sample}
We calculated numerical solutions for initial black hole masses $M$ ranging 
from $2^{-10}$ to $16$ (\ref{initial_cond}). 
Here, we present the results for $M=8$ as the macroscopic case for uniformity of exposition.
All cases show similar convergence behavior for the regularized fields, though as we approach $M=16$, 
derived physical quantities start to show irregular convergence patterns due to catastrophic cancellation.
The fact that $M=8$ is sufficiently large to be categorized as ``macroscopic'' will be established elsewhere, 
when we discuss the physical interpretation of our results.

The regularized fields $\bar{\theta}$ and $\bar{\phi}$ from solutions with two
values of $M \gg 1$ and $M \ll 1$ are shown in Figs'~\ref{M8N24_phireal} and \ref{M9.765625e-04N24_phireal}. 
As discussed before, a central issue with the numerical calculations is to ensure 
that we get close to the last ray and the singularity, as many of the interesting 
phenomena occur in this region. It is analytically known that the singularity 
of the CGHS model occurs when $\Phi = \frac{N}{12}$. Moreover, $\Phi - \frac{N}{12}$ evaluated on the dynamical horizon 
(determined by $\partial_+ \Phi =0$) can be interpreted as the quantum corrected area of the black hole 
\cite{PiranStrominger}. This way, we can test our proximity to the singularity by checking the value of the area near the 
singularity---see Fig.\ref{M16N24_Area}. For $M \gg 1$, the part of our compactification scheme which emphasizes 
the region near the last ray is crucial to reach the region where the black hole area drops to less than 
{\em a few percent} of its initial value, let alone to near the Planck mass. 
As explicitly seen in the figure, had we used
a uniform mesh in uncompactified $z^{\pm}$ coordinates, a mesh spacing of order $h \tilde{<} 10^{-M}$
would have been needed. Covering a sufficient region of the spacetime to reveal
the asymptotics would require a net coordinate range $\Delta z^\pm$ of order unity, implying
a mesh of order $10^{M}$ points along both directions, which is of course impractical to achieve
on contemporary computers for larger $M$. This important aspect of the problem was not clear in earlier
studies, as they usually focused on $M \sim 1$ 
\cite{PiranStrominger,Lowe}. As we will describe in the companion paper on the
physical results, the $M< 1$ solutions are drastically different from the $M>1$ solutions.

A final comment---even with compactification, eventually finite precision floating
point arithmetic will limit how large an initial mass we can simulate; with long double
precision (80-bit), we are restricted to $M<\sim20$.
%
%
\begin{figure}
\includegraphics[scale=.85]{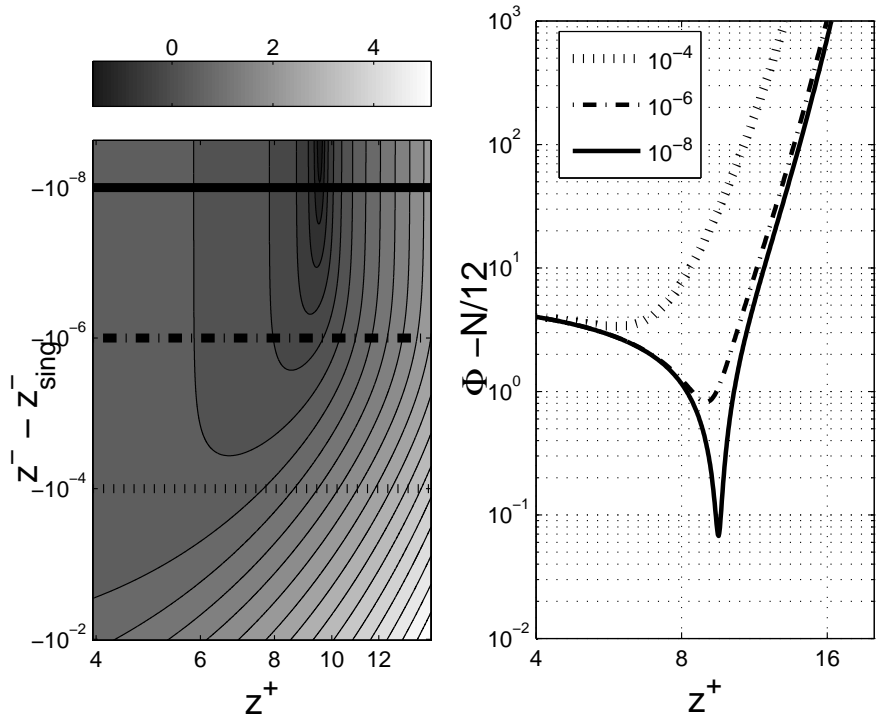}
\caption{ $\Phi$ for $M=8$, $N=24$. Left: Base-$10$ logarithm of $\Phi-\frac{N}{12}$. Right: $\Phi-\frac{N}{12}$ at lines of constant $z^-_{sing}-z^- = 10^{-4}, 10^{-6},10^{-8}$. This shows that $\Phi$ approaches $N/12$, the location of the spacetime
singularity, from where the last ray eminates. Specifically, here $\Delta z^- \sim 10^{-8}$ of the last ray. } 
\label{M8N24_phireal}
\end{figure}
%
%
\begin{figure}
\includegraphics[scale=0.92]{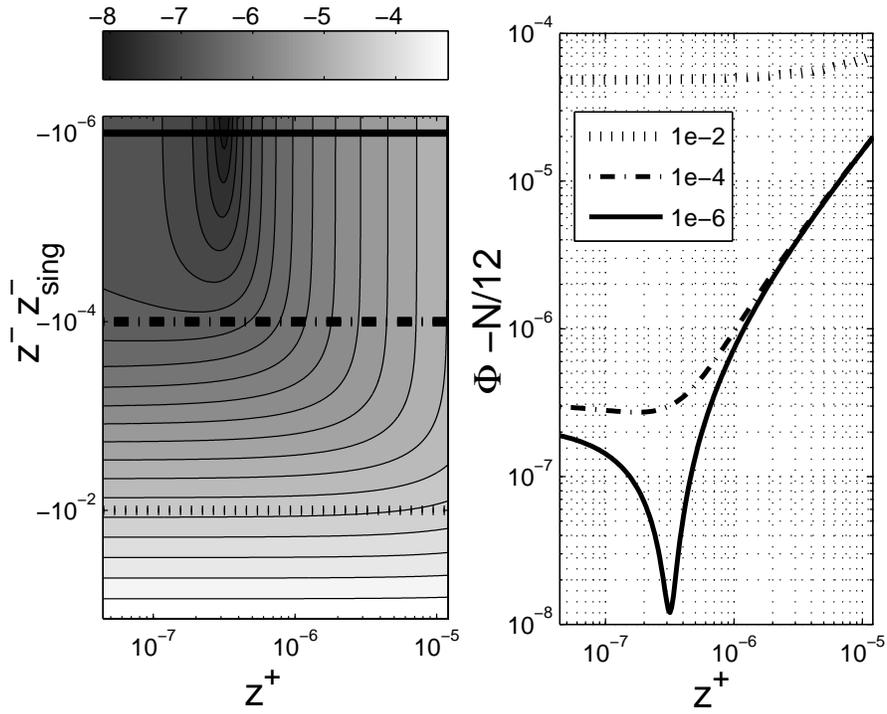}
\caption{ $\Phi$ for $M=2^{-10}$, $N=24$. Left: Base-$10$ logarithm of $\Phi-\frac{N}{12}$ Right: $\Phi-\frac{N}{12}$ at lines of constant $z^-_{sing} - z^-= 10^{-2}, 10^{-4},10^{-6}$. Again, as in Fig.~\ref{M8N24_phireal}, this shows that $\Phi$ approaches $N/12$,
and we are close to the location of the last ray.
Note that the field values are generally quite different from the $M=8$ case, and the singularity appears very close to $z^+ =0$, which necessitated the special compactification scheme explained in Sec.~\ref{sec_compact}. } 
\label{M9.765625e-04N24_phireal}
\end{figure}
%
%

\begin{figure}
\includegraphics[scale=0.88]{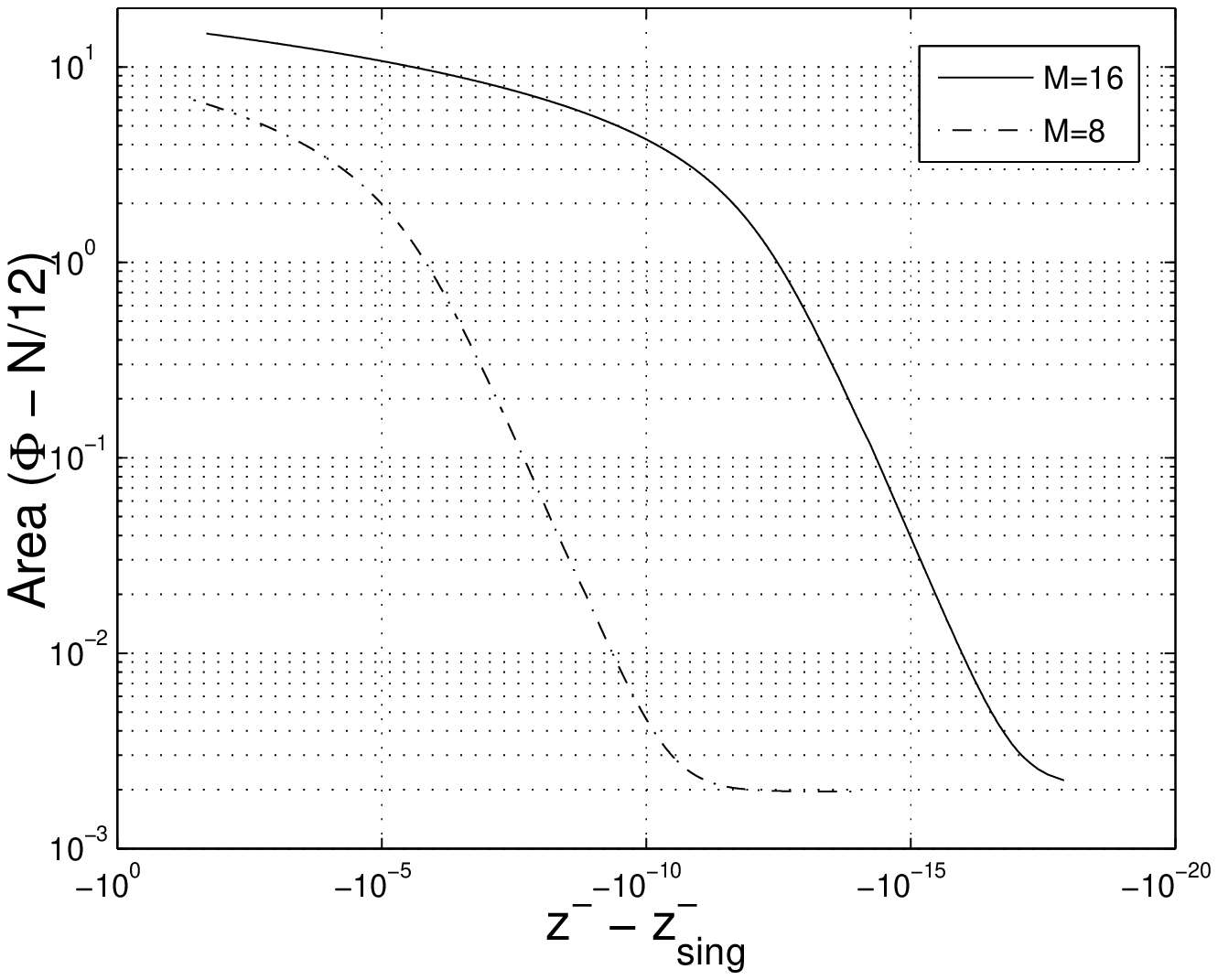}
\caption{ Area of the black hole ($\Phi-\frac{N}{12}$) vs. the uncompactified distance from the last ray in a log-log plot for 
$M=8,16$ and $N=24$. Note that in terms of the uncompactified coordinates, we have to be within $\Delta z^- \sim 10^{-8}$ of the 
last ray in order to be truly close to the singularity for $M=8$, and within $\Delta z^- \sim 10^{-16}$ for $M=16$. This exponential trend is general and severely limits the upper value of $M$ we can use in numerical calculations if we want to reach regions ``close'' to the singularity.} 
\label{M16N24_Area}
\end{figure}

\subsection{Convergence of the Fields}
We compute convergence factors by comparing solutions obtained using different 
mesh spacings. Note that we are using the Richardson extrapolation scheme
described in Sec.~\ref{sec_prep}, thus in the following when we refer to a solution 
computed with mesh spacing $h$, $h$ labels the coarsest resolution mesh of the
four used in the numerical integration.

First, we define 
\be
\Delta_h f \equiv f_{h}-f_{h/2} \ 
\ee
where $f_{h}$ denotes the numerical solution of a function $f$ obtained
on a grid with mesh spacing $h$. $\Delta_h f$ is thus an estimate, to $O(h^{n})$, 
of the truncation error in $f$, where $n$ is the rate of convergence of the algorithm.
From the Richardson expansion we then get
\be
n = \log_2 \left[\frac{\Delta_{2h} f}{\Delta_h f} + O(h)\right] 
  = \log_2 \left[\frac{f_{2h}-f_{h}}{f_{h}-f_{h/2}} + O(h)\right],
\ee
where the next-to-leading order term is of $O(h)$ because
of the order of interpolating polynomial we use. From the above,
we define an estimated convergence factor $n_e$ via
\be
n_e\equiv \log_2 \frac{f_{2h}-f_{h}}{f_{h}-f_{h/2}}
\label{n_conv}
\ee
\begin{figure}
\includegraphics[scale=0.88]{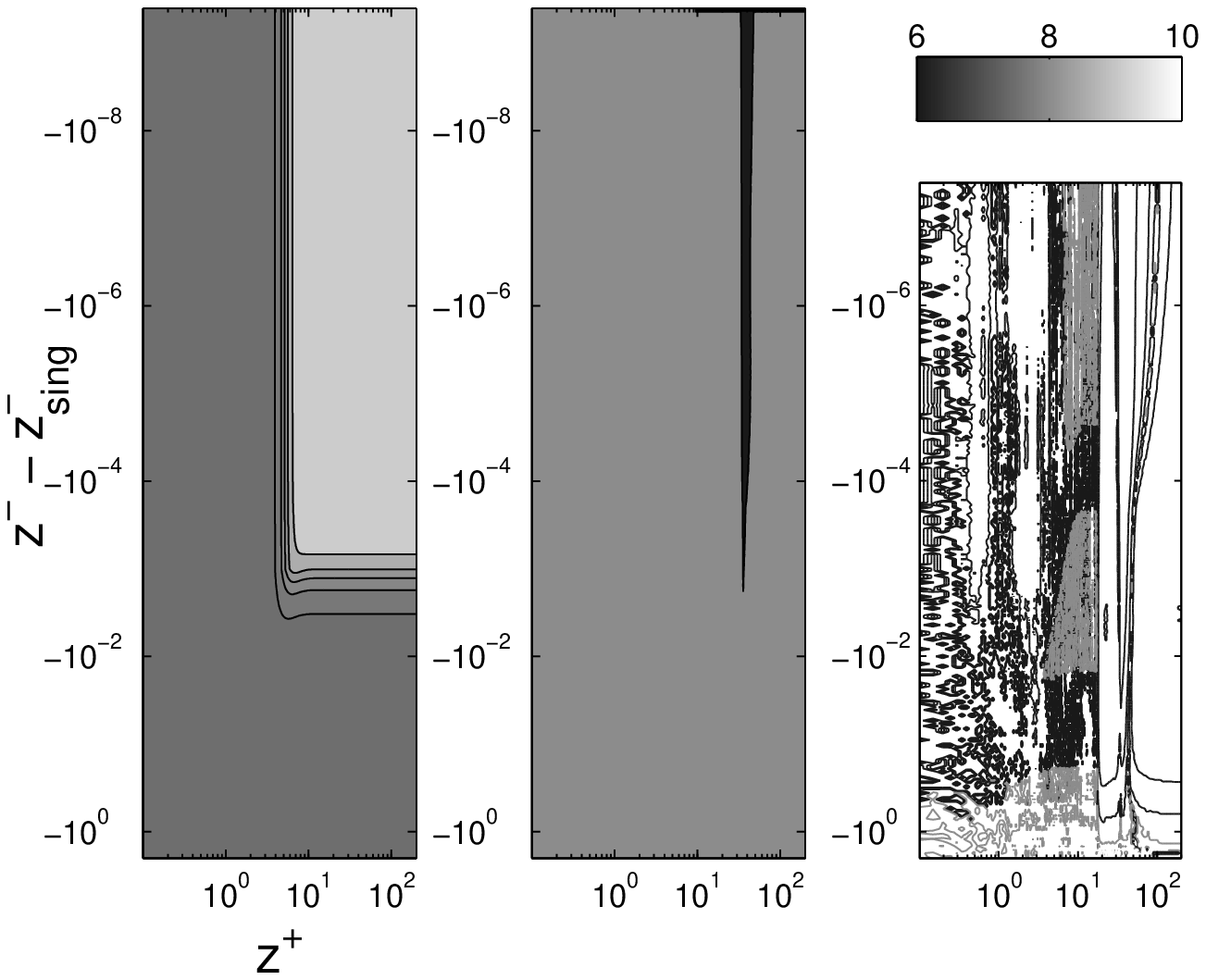}
\caption{ Convergence of the $M=8$, $N=24$ case: $n_e(z^{\pm})$ for $h=2^{-10}$ (left) is mostly in the range $9-10$, and
for $h=2^{-11}$ (middle) is around $8$. For $h=2^{-12}$ (right) we reach machine round-off, and thus
loose convergence, hence the ``noisy'' pattern.}
\label{M8N24n_e_phi}
\end{figure}
\begin{figure}
\includegraphics[scale=0.88]{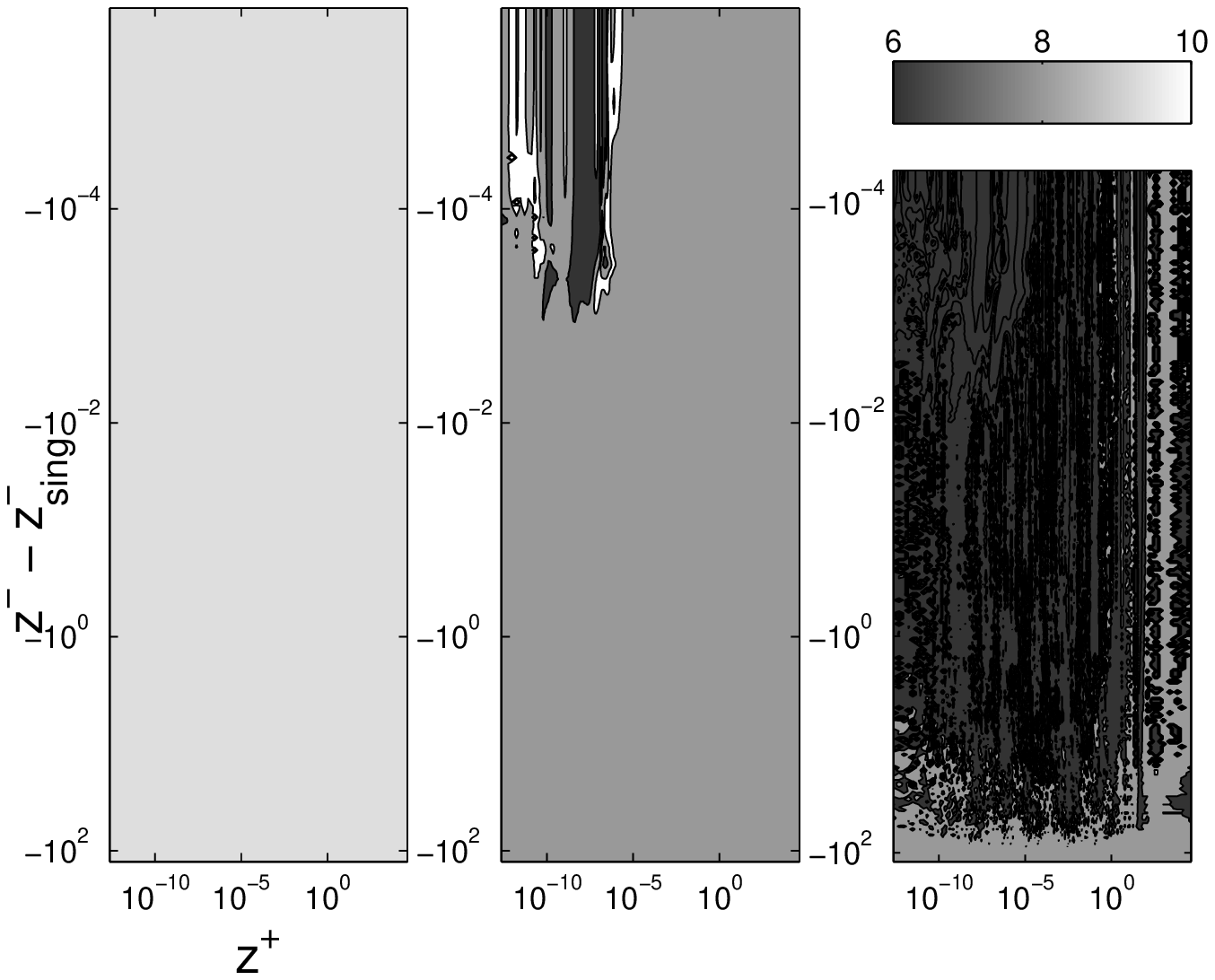}
\caption{ Convergence of the $M=2^{-10}$, $N=24$ case: $n_e(z^{\pm})$ for $h=2^{-10}$ (left) is around $10$, and for $h=2^{-11}$ (middle) is around $8$. Again, as with the $M=8$ case in Fig.~\ref{M8N24n_e_phi}, for $h=2^{-12}$ (right) machine round-off error
begins to dominate the error, hence the ``noisy'' pattern. 
This effect is already visible in certain regions of the $h=2^{-11}$ case. For lower mass black holes, round-off is reached with coarser meshes relative to the higher mass black holes.}
\label{M9.765625e-04N24n_e_phi}
\end{figure}
In Figs.~\ref{M8N24n_e_phi} and \ref{M9.765625e-04N24n_e_phi} we show plots
of $n_e$ for a high and low mass case respectively. An ``issue'' we have with the 
convergence behavior of the CGHS equations is it seems artificially high for coarser
meshes. One reason for this may be that the 
central difference scheme (\ref{eq:difference_equations}) we use solves the homogeneous 
part of the wave equation ($\partial_+ \partial_- f=0 $) {\em exactly} (to within round-off), 
irrespective of the step size. Furthermore, with our choice of variables and regularization
scheme, it is only the non-linear quantum corrections that introduce non-trivial evolution,
and initially the effects of this will be small. Though regardless, 
in the limit of zero $h$ we should approach the expected convergence
behavior; as shown in these figures, we {\em do} see this trend, though have not
quite reached the limiting behavior before machine round-off error is reached.

As mentioned, reasons for the anomalous convergence behavior may be the compactification
and special initial data we choose, namely regularized fields that are initially
adapted to the classical solution. 
To check this, we evolved a test case where we imposed the initial conditions 
for $M=11$, $N=11$ at $z^+_c =0.25$ rather than $\scri^-_R$. Note that this
is not a physically correct solution as it will violate the constraints, though
it is mathematically perfectly valid non-trivial initial data for the evolution equations.
We set the domain of computation to  $z_c^- \in [0.25,0.5]$ and $z_c^+ \in [0.0,0.25]$ to avoid any singular behavior. 
Using four meshes for the Richardson extrapolation in this test, the truncation error was 
again reduced down to round-off level for even the coarser meshes, so for this test alone,
we only employed three successively finer meshes in the extrapolation scheme; 
hence the resulting truncation error is expected to scale as $h^6$. The result
is shown in Fig.~\ref{M11N11n_e_phi_temp}, where we see the expected convergence.
We also tested two-mesh Richardson extrapolation for the same case,
and obtained the expected $h^4$ convergence.

\begin{figure}
\includegraphics[scale=0.90]{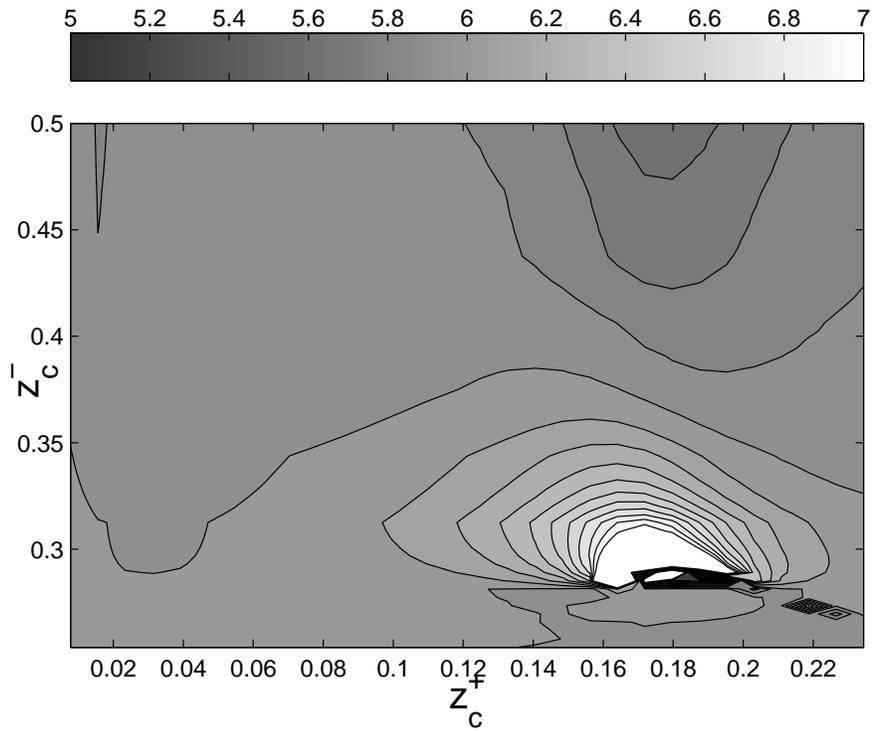}
\caption{The convergence factor $n_e$ for $h=2^{-8}$ where as a test we imposed the (unphysical) initial conditions 
for $M=11$, $N=11$ at $z^+_c =0.25$ rather than $\scri^-_R$. 
We only evolved the fields in the region $z_c^- \in [0.25,0.5], z_c^+ \in [0.0,0.25]$. 
This solution is not physically relevant, though tests the
behavior of the numerical code away from any of the null infinities or singularities.
Here, for each base resolution, three meshes where used in the Richardson 
extrapolation scheme, which should give $O(h^6)$ convergence, and does to good approximation
as shown in the figure.}
\label{M11N11n_e_phi_temp}
\end{figure}

\subsection{Convergence of Physical Quantities on $\scri_R^+$ }

The physical quantities we are interested in, including $y^-(z^-)$, $F_{ATV}$ and $M_B$, are
all functions of the fields, thus in theory they should inherit the convergence behavior
of the fields. Some of these quantities require computing first and
second derivatives of the fields, and so to maintain the theoretical convergence factor of $7$,
one should use $9$-point finite difference stencils. However, catastrophic cancellation plagues the numerical derivatives
near the last ray, as the regularized fields vary extremely slowly in this region, and this seems to be the limiting factor in 
the accuracy in which we can compute physical quantities. 
Though in general we do not need high order convergence of derived quantities to achieve
high accuracy. A case-and-point is $M_B$, obtained by integrating $F_{ATV}$. $F_{ATV}$ is dominated by round-off near the last ray 
in most cases, though, once integrated over $y^-$, this region contributes insignificantly to $M_B$. 
Furthermore, simple trapezoidal integration is adequate to achieve quite accurate estimates
of $M_B$, as illustrated in Fig.~\ref{M8and9.765625e-04N24n_truncation_M_B}. 

\begin{figure}
\includegraphics[scale=0.88]{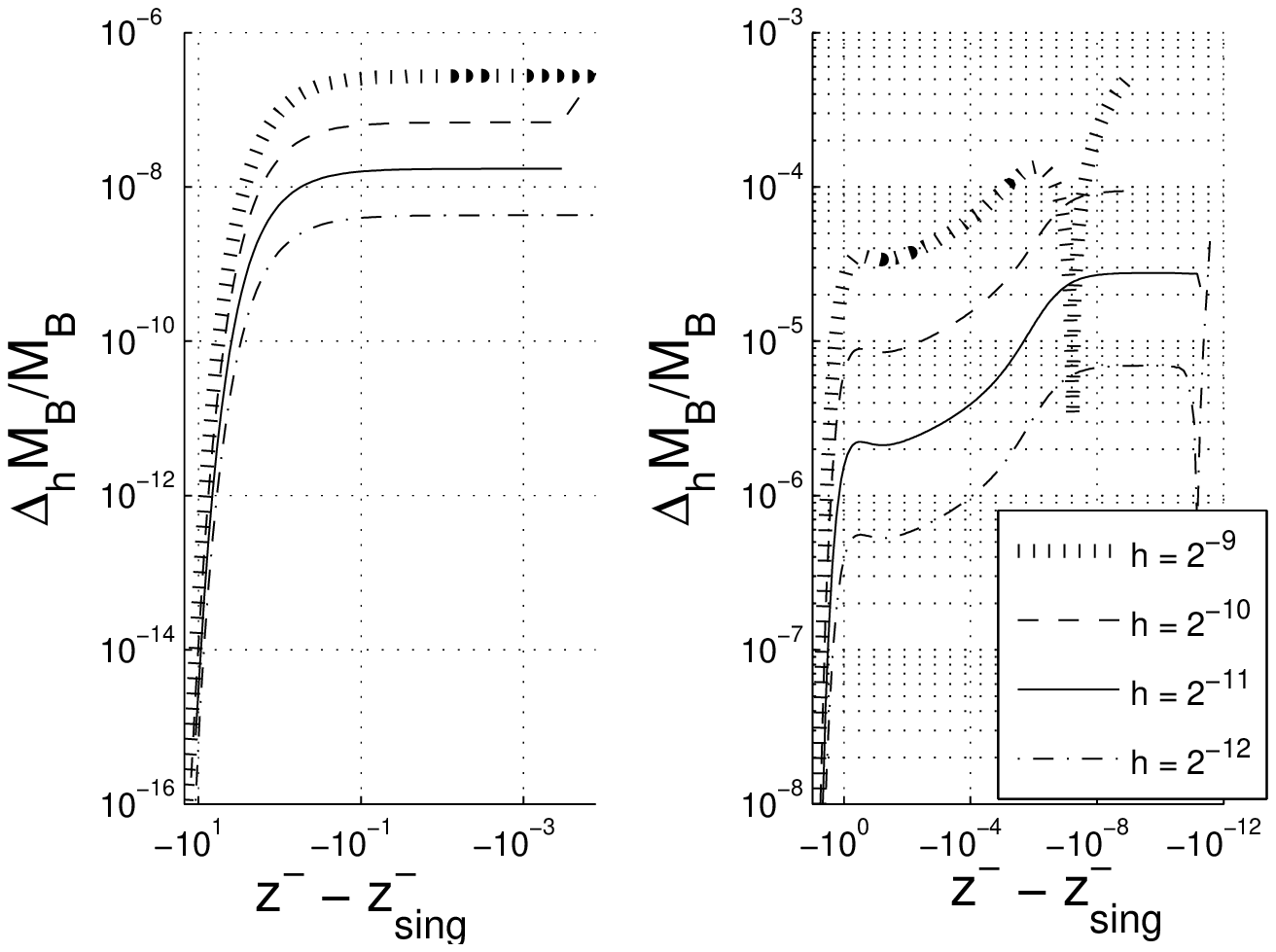}
\caption{$ \left| \frac{\Delta_h M_B}{M_B} \right| $ for various values of $h$ for 
$M=2^{-10}$ (left) and $M=8$ (right). For most of the range, there is clear quadratic convergence. The dominant
error here is from the trapezoidal integration method, and not a reflection of the truncation error from the numerical 
calculation of the fields.}
\label{M8and9.765625e-04N24n_truncation_M_B}
\end{figure}

\subsection{Independent Residuals}

As a final test of the code, from numerical solutions, we compute independent
residuals of the differential equations (\ref{eq:evolution_regularized1}) and
(\ref{eq:evolution_regularized2}). Specifically, we calculate the derivatives 
using three-point stencils centered at the mesh points, rather than the cell centered differences
used for the solution. Three-
point stencils limit the convergence of the independent residual to quadratic order, regardless of the 
convergence of the fields themselves. We observe the expected quadratic convergence in all cases.

\section{Conclusions}\label{sec_conc}
In this paper we presented some details of the technical aspects of our numerical solution of the CGHS model.
Beside the links this model provides to black hole evolution in $3+1$ dimensions
(which will be published elsewhere), it also presents considerable numerical challenges to solve, 
in particular in the macroscopic mass limit.
The fact that the CGHS model has two distinct regimes in parameter space, $M \gg N/24$ and 
$M \ll N/24$, has not been emphasized in the literature before. Not only do these two regimes have 
radically different physical properties and interpretations, their numerical analysis also presents 
considerably different levels of challenge. Existing numerical studies of the CGHS model focused on the 
intermediate mass range $M \sim N/24$, for example $\frac{M}{24N}=1$ in~\cite{PiranStrominger} and $\frac{M}{24N}=2.5$ in~\cite{Lowe}. This case, like the low mass region, is considerably 
easier to solve numerically. In this regime, most of the evolution of the fields is not confined to 
a small region near the last ray, thus, the uncompactified $z^{\pm}$ coordinates are adequate
to cover the quantum-corrected spacetime. However, many of the interesting phenomena of the black hole evaporation cannot be observed in this regime. In the physically more interesting case of macroscopic black holes, we are led to using compactified coordinates: we need to start the 
calculation sufficiently far away from the last ray, yet at the same time have high 
resolution ($\Delta z^- \sim 10^{- 16}$ for $M=16$, $N=24$) near the last ray.

A correct estimate of the position of the singularity necessitates very low truncation error, which 
we obtain using Richardson extrapolation (with intermittent error removal). 
This takes the results from four successively finer meshes to 
obtain results that theoretically scale as $\mathcal{O}(h^7)$, and we use this scheme to reduce the truncation 
error to the level of machine precision ($10^{-16}$). 
Even though there is still space for improvement of our method using improved
compactification schemes, higher precision numerics, etc.
we have enough accuracy to, for the first time, discern the physics of the CGHS model near the last 
ray  in the macroscopic mass regime.

\textbf{Acknowledgments:} We would like to thank Abhay Ashtekar who suggested the problem to us, collaborated with us on the physical aspects of the CGHS model, and made useful
suggestions on this manuscript. We would also like to thank Amos Ori for
useful discussions. We acknowledge support from NSF grant PHY-0745779,
and the Alfred P. Sloan Foundation (FP). Some of the simulations were
run on the {\bf Woodhen} cluster at Princeton University.

\end{document}